\documentclass[conference]{IEEEtran}
\IEEEoverridecommandlockouts
\usepackage{cite}
\usepackage{booktabs} 
\usepackage{caption}
\usepackage{graphicx, subfig}
\usepackage{amsmath,amssymb,amsfonts}
\usepackage{graphicx}
\usepackage{textcomp}
\usepackage{xcolor}
\usepackage{multirow}
\usepackage{ragged2e} 
\usepackage{booktabs,makecell, multirow, tabularx}
\usepackage{tabularx}
\usepackage{makecell}
\usepackage{multicol}
\usepackage{paralist,bbding,pifont}
\usepackage[ruled]{algorithm2e} 
\usepackage{dsfont}
\begin{document}

\title{Robust Representation Learning for Unified Online Top-K Recommendation
}

\author{
\IEEEauthorblockN{Minfang Lu\IEEEauthorrefmark{1}\IEEEauthorrefmark{3},
Yuchen Jiang\IEEEauthorrefmark{1}\IEEEauthorrefmark{2}\thanks{\IEEEauthorrefmark{1}Equal contribution}, Huihui Dong\IEEEauthorrefmark{2}, Qi Li\IEEEauthorrefmark{2}, Ziru Xu\IEEEauthorrefmark{2}, Yuanlin Liu\IEEEauthorrefmark{3},\\Lixia Wu\IEEEauthorrefmark{3}, Haoyuan Hu\IEEEauthorrefmark{3}, Han Zhu\IEEEauthorrefmark{2}, Yuning Jiang\IEEEauthorrefmark{2}, Jian Xu\IEEEauthorrefmark{2}, Bo Zheng\IEEEauthorrefmark{2}}
\IEEEauthorblockA{\textit{Alibaba Group}\IEEEauthorrefmark{2},\textit{Cainiao Network}\IEEEauthorrefmark{3}}
\IEEEauthorblockA{\textit{Beijing, China\IEEEauthorrefmark{2}, Hangzhou, China\IEEEauthorrefmark{3}}}
\IEEEauthorblockA{\{luminfang.lmf, jiangyuchen.jyc, dhh267344, luyuan.lq, ziru.xzr, yuanlin.lyl\}@alibaba-inc.com \\ \{wallace.well, 
 haoyuan.huhy, zhuhan.zh, mengzhu.jyn, xiyu.xj, bozheng\}@alibaba-inc.com}
}

\maketitle
\begin{abstract}
In large-scale industrial e-commerce, the efficiency of an online recommendation system is crucial in delivering highly relevant item/content advertising entities that cater to diverse business scenarios. However, most existing studies focus solely on item advertising, neglecting the significance of content advertising. This oversight results in inconsistencies within the multi-entity data structure and unfair retrieval of different entities within the recommendation system. Furthermore, the challenge of retrieving top-k advertisements from multi-entity advertisements across different domains adds to the complexity. Recent research proves that user-entity behaviors within different domains exhibit characteristics of differentiation and homogeneity. Therefore, the multi-domain matching models typically rely on the hybrid-experts framework with domain-invariant and domain-specific representations. Unfortunately, most approaches primarily focus on optimizing the combination mode of different experts, failing to address the inherent difficulty in optimizing the expert modules themselves. The existence of redundant information across different domains introduces interference and competition among experts, while the distinct learning objectives of each domain lead to varying optimization challenges among experts. To tackle these issues, we propose robust representation learning for the unified online top-k recommendation. Our approach constructs unified modeling in entity space to ensure data fairness. The robust representation learning employs domain adversarial learning and multi-view wasserstein distribution learning to learn robust representations. Moreover, the proposed method balances conflicting objectives through the homoscedastic uncertainty weights and orthogonality constraints. Various experiments validate the effectiveness and rationality of our proposed method, which has been successfully deployed online to serve real business scenarios.
\end{abstract}


\begin{figure}[htbp]
\centering
\includegraphics[scale=0.25]{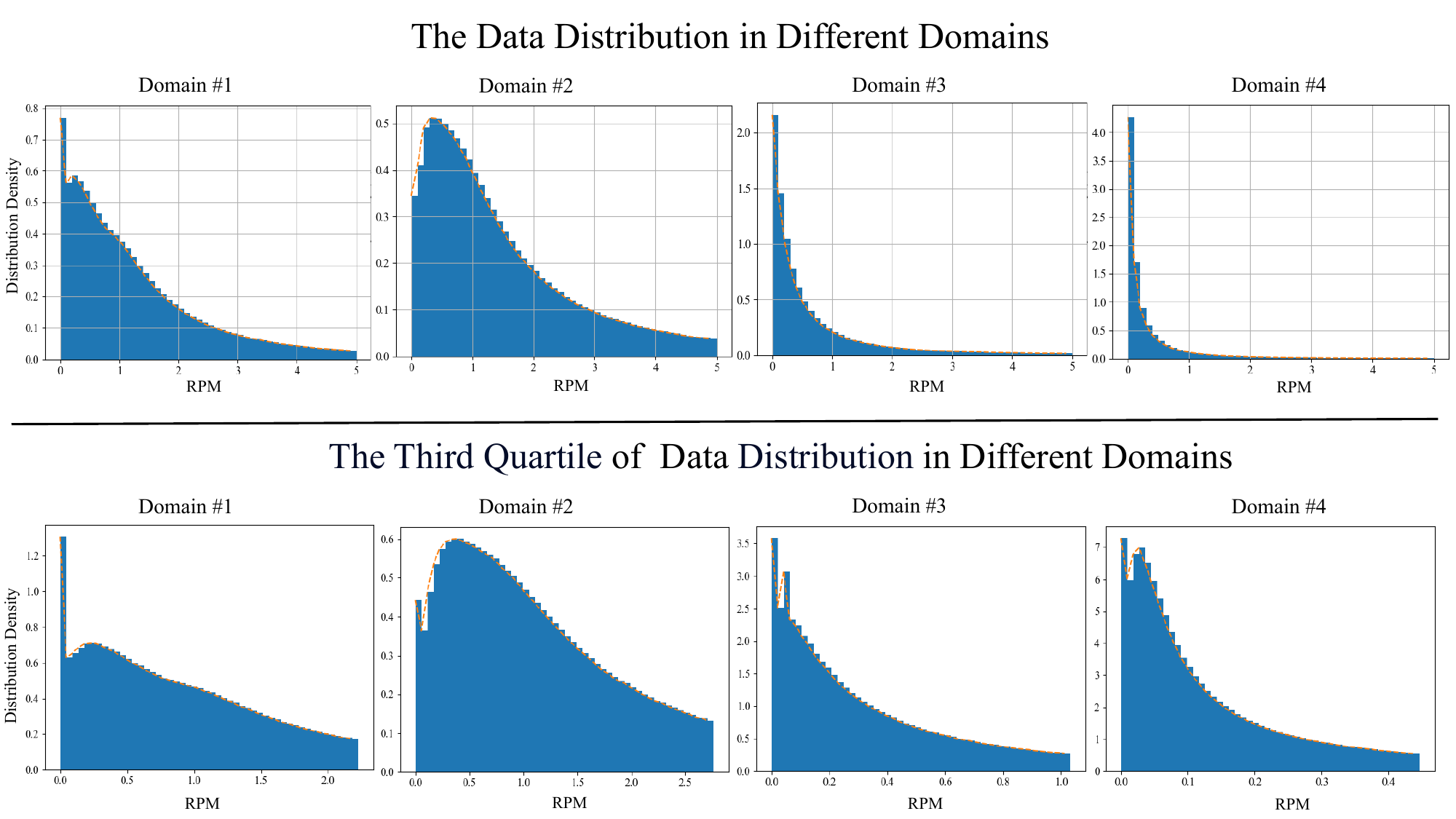}
\caption{The first line shows the normalized distribution of the metric RPM corresponding to all data in different domains. The second line shows the normalized distribution of the metric RPM corresponding to quantile-75 of data in different domains. In each subgraph, the abscissa is RPM, and the ordinate is the distribution density.}
\label{fig_data_distribution}
\end{figure}

\section{Introduction}
The recommendation system~(RS) often consists of different modules: matching, ranking, reranking, and so on. 
As the first stage of the whole RS, the matching module needs to retrieve top-k highly relevant items for each user from hundreds of millions of candidates, which determines the upper limit of the RS. 
Collaborative filtering~(CF)\cite{shi2014collaborative} is a basic algorithm of personalized RS and is still widely used in many industrial scenarios. 
CF calculates the similarity score by the dot product of low-dimensional user and item representations and then predicts user interest in the whole candidates. 
However, the data sparsity issue severely limits the expressive ability of CF, as users consume only a small number of entities across the entire item/content space\cite{xu2018deep}.

With the wide application of deep learning, the deep matching methods\cite{GoogleDNN,TDM,GRU4Rec} achieve both accurate and relevant performance by expressing the matching function through the deep learning model and matching with the high-performance approximate retrieval algorithm. The deep matching learning model can capture higher-order feature interactions in sparse and multi-modal data in an end-to-end manner. In a single business domain, the deep matching model of the RS only needs to consider the user's consumption behavior in the current domain. 
However, in large-scale industrial e-commerce, an efficient online RS needs to deliver multiple advertising entities tailored to diverse business scenarios. 
For instance, Alibaba's e-commerce platform has various advertising entities, such as item and content advertisements~(ads). Content ads encompass a range of entities with significant structural differences, such as shops, short videos, and live broadcasts. These entities are placed across various business scenarios, such as the home page, the shopping cart page, and the mini-game pages to maximize user engagement and facilitate browsing experiences. 

When confronted with vast amounts of multi-entity advertising data, constructing a suitable feature system for the RS and modeling advertising representation have become crucial considerations. However, most existing studies tend to concentrate solely on item ads, disregarding the importance of content ads. The inconsistency between offline/online training based on multi-entity data links leads to the unfair retrieval of different entities in the RS.
Simultaneously, multi-entity ads within different domains may exhibit varying levels of abundance or scarcity, while user behaviors may be singular or diverse. 
The data distribution in Alibaba's display advertising online log on a certain day is shown in Figure \ref{fig_data_distribution}.
The current multi-entity and multi-domain advertising landscape presents significant challenges for contemporary online RS.
A potential solution is designing distinct matching models for each entity and domain. However, this approach is unfeasible for real-world applications due to its high computational costs, repetitive development, and data silos phenomenon \cite{AdaSparse}.

The other solution aggregates samples from various entities and domains into a single dataset to train a single-domain deep matching model. However, this approach presents challenges:
1). Since different entities have distinct feature construction, constructing a suitable unified item/content underlying advertising features becomes crucial to ensure fairness across different entities.
2). The deep matching model needs to score entities with different feature distributions and data volumes in a unified manner, which requires accurately extracting representations of multi-entity ads. 
3). While aiming for a unified modeling approach across multiple entities, a single-domain deep matching model may fail to capture the unique information associated with each domain\cite{MDLwithCWPC}.
Given the challenges above, a question arises here: \textbf{can we design an excellent online matching module that can comprehensively consider multi-entity advertising data modeling and multi-domain distribution modeling to adapt to the differences of various domains/entities flexibly?} 

Regarding multi-entity modeling, it becomes imperative to establish a unified data feature link for multi-entity ads, which ensures feature consistency between various entities. Furthermore, achieving fair retrieval of item/content ads relies on the matching model's unified representation ability. 
Additionally, from the deep multi-domain matching learning perspective, it is necessary to model the commonality and differences of the domains with different data distributions, which can improve the model performance for each domain under the premise of controllable costs.
To our knowledge, this work is the first attempt at the unified modeling of the multi-entity and multi-domain problem.
Research\cite{STAR, chen2023cold, niu2021heterogeneous} in the multi-domain model for the RS has primarily focused on the ranking stage, disregarding the potentially even more critical matching stage. By shifting the methodology to the matching stage, the matching model can better comprehend the intricate relationships between users and items across multiple domains\cite{zhang2022multi}. The multi-domain matching model facilitates the retrieval of more diverse and relevant items, provides a wide-ranging item set for subsequent stages, and ultimately significantly improves the overall performance and efficacy of the RS. 

The ADIN\cite{ADI} and MDNN\cite{MDNN} have validated the effectiveness of the hybrid expert framework in multi-domain matching learning. They model shared-domain experts for domain-invariant representation and specific-domain experts for specific-domain representation, respectively. While these methods have focused on optimizing the combination mode of different experts, they have yet to notice the inherent difficulty in optimizing the diverse experts themselves. The distinct learning objectives in each domain pose unique optimization challenges for the experts. On the one hand, domain-invariant representations must suppress domain-specific noise while capturing general features. However, different domains exhibit diverse distributions and invariants, making learning a uniform and effective representation across all domains difficult. On the other hand, learning specific-domain representations focuses on personalized modeling within each domain to capture specific information. Furthermore, those experts share specific parameters during training in the hybrid-experts framework, which can introduce interference and competition. The delicate trade-off between learning domain-invariant and domain-specific representations becomes further complicated.

In summary, the main contributions are listed below:
\begin{itemize}
  \item [1)]We innovatively propose a unified online top-k recommendation method. It integrates the item and content ads through the unified modeling in entity space and multi-domain matching model, achieving high reusability and minimizing development costs. The unified multi-entity link, ensuring the fairness of retrieval for different entities, decouples with the multi-domain matching model and remains compatible with various model structures. Notably, this work represents the pioneering effort in RS toward multi-entity and multi-domain unified modeling.
  \item [2)]We present a novel robust representation learning(RRL) for the multi-domain matching model. RRL utilizes multi-view wasserstein discriminators to facilitate the domain-invariant representation for improved generalization and employs domain adversarial learning to maximize the multi-domain discrepancy for domain-specific representations. Furthermore, RRL achieves the balance between conflicting objectives by the homoscedastic uncertainty weight and orthogonality constraint.
  \item [3)]Our various experiments demonstrate the validity and rationality of the proposed method. In addition, the unified matching module with RRL is deployed online, serving real business scenarios within Taobao and Cainiao. It improves the online retrieving ability of different entities, unifies the online redundant logic, and facilitates the expansion of new entities and domains.
\end{itemize}

\section{Related Work}
\subsection{Online Recommendation system}
The proliferation of the Internet and smart devices has led to the widespread popularity and convenience of e-commerce. Major online shopping platforms like Alibaba and Amazon offer various products. The primary objective of RS is to suggest suitable products to users accurately. RS has formed stable recommendation paradigms consisting of matching, ranking, and reranking, integrating knowledge acquisition, machine learning, and text classification, among other technologies \cite{hussien2021recommendation,eliyas2022recommendation,xia2022multi}. However, as application scenarios grow increasingly complex, there is a growing need to efficiently and cost-effectively deploy new recommendation methods that can quickly adapt to diverse business scenarios and accommodate multi-entity ads. MTMS\cite{tan2021multi} explores independent/non-shared embeddings for each task and domain in the RS, adapting new domains conveniently by reducing the coupling between tasks and scenarios. The similar works that focused on the multi-task combined with the multi-domain model include HiNet\cite{zhou2023hinet} and M2M\cite{zhang2022leaving}. DeepIDRS\cite{islek2022hierarchical} improves performance by designing a hierarchical interpretable RS that accurately represents subsets of project titles, descriptions, and project comments. Some RS research works have focused on content recommendations, particularly video\cite{wei2019mmgcn}, audio\cite{deldjoo2020recommender}. However, there are various entities of online content recommendation. Developing separate models for each specific content type would be challenging. In fact, there are also commonalities and distinctions in item/content entities. By incorporating a unified multi-entity feature structure in data space, RS can leverage cross-entity information to enhance the recommendation of content ads with scarce data.
\subsection{Deep Matching Learning}
With the powerful representation learning capability of deep learning and the rapid growth of computing power, deep matching learning has recently made significant progress, which can be categorized into representation learning methods and matching function learning methods\cite{xu2018deep}. Representation learning methods focus on user and item representations to generate matching scores, such as sequential interactions\cite{GRU4Rec}, multi-modal content\cite{ACF}, and graph data\cite{GraphRec}. Matching function learning methods emphasize the interaction between users, items, and context to generate matching signals, such as implicit interaction modeling\cite{deepcrossing}, explicit interaction modeling\cite{HoAFM}, and the combination of explicit and implicit interaction modeling\cite{xdeepfm}. 
However, these works are mainly designed for one particular business domain. The unified feature embeddings learned from a global feature space are served across different business domains. With the increasing complexity of industrial scenarios, conventional deep-matching learning methods are gradually becoming insufficient to handle large-scale RSs. As a result, multi-domain deep matching learning has emerged as a novel approach that emphasizes learning domain-invariant and domain-specific representations, which solve the bottleneck for adequate characterization and interaction in complex industrial scenarios.

\subsection{Multi-Domain Learning}
Multi-domain learning involves sharing information about the same problem across different contextual domains with different data distributions. On the contrary, multi-task learning involves sharing information about different problems within the same domain with the same data distribution. There has been numerous research\cite{MDLMTL1,MDLMTL2,MDLMTL3} highlighting the differences between these two approaches. In the RS, multi-domain learning is typically used for ranking and matching phases. HMoE \cite{HMoE} exploits business relationships in the label space by utilizing a stacked approach based on MMOE \cite{mtlMMoE}. STAR \cite{STAR} proposed a star topology consisting of shared centered parameters and domain-specific parameters for explicitly learning commonalities and distinctions between different domains. MGFN\cite{zhang2022multi} leverages a graph convolution module to capture domain-specific patterns, combining a fusion module to obtain a global representation and utilizing a specific cross-domain negative sampling strategy. ADIN\cite{ADI} adopts a shared-specific hybrid-experts framework with self-learning, including the domain interest adaptive layers for user and item towers. MDNN\cite{MDNN} designs an anomalous multi-branch network based on deep neural networks for learning latent knowledge. 
While many existing multi-domain matching methods focus on optimizing the hybrid-experts combination and extracting user interests in multiple domains, they do not pay attention to the inherent learning difficulties in experts themselves and neglect the trade-off between the contradictory representations of the shared-domain and specific-domain experts. Therefore, we propose robust representation learning for the multi-domain matching model based on domain adversarial learning and multi-view wasserstein distribution learning, which can adapt to different domains and entities and allow the model to transfer the representation to new domains or entities quickly.

\subsection{Adversarial Domain Adaptation} 
Domain adaptation is a common technique in transfer learning, where knowledge is learned from one or more source domains with sufficient labeled data for use in a target domain with insufficient data. Adversarial learning\cite{GAN} is to exercise the distributed learning ability through the minimax game to obtain generated data that can confuse real data. DANN\cite{DANN} innovatively proposes to use adversarial learning to extract distinguishable features of labeled source data and indistinguishable features of source and target domains to align distributions.
DAAN\cite{DAAN} learns domain-invariant representations dynamically while proposing dynamic adversarial factors to evaluate the relative importance of marginal and conditional distributions.  
NWD\cite{NWD} reuses the original task-specific classifier by coupling it with the nuclear-norm wasserstein discrepancy to construct a discriminator that satisfies the K-Lipschitz constraints implicitly. 
Based on the previous work, we propose the RRL for the multi-domain matching model, which aims to address the challenges arising from different optimization goals among experts. RRL obtains invariant/specific representation from shared-domain/specific-domain experts with domain invariant adversarial learning based on the gradient reversal layer. In contrast to previous work, we design multi-view wasserstein distribution learning for shared-domain experts, which explicitly aligns the distribution of domain-invariant representations across different domains, enhancing the universality of the domain-invariant representations. Additionally, RRL introduces orthogonal constraints among the experts to minimize the interference of redundant information between shared-domain and specific-domain experts, ensuring that each expert focuses on its objectives without being influenced by irrelevant information. To balance the trade-off between contradictory objectives during training, the RRL utilizes homoscedastic uncertainty weight, which adaptively adjusts the weights of different constrained objectives.

\begin{figure*}[htbp]
\centering
\includegraphics[scale=0.15]{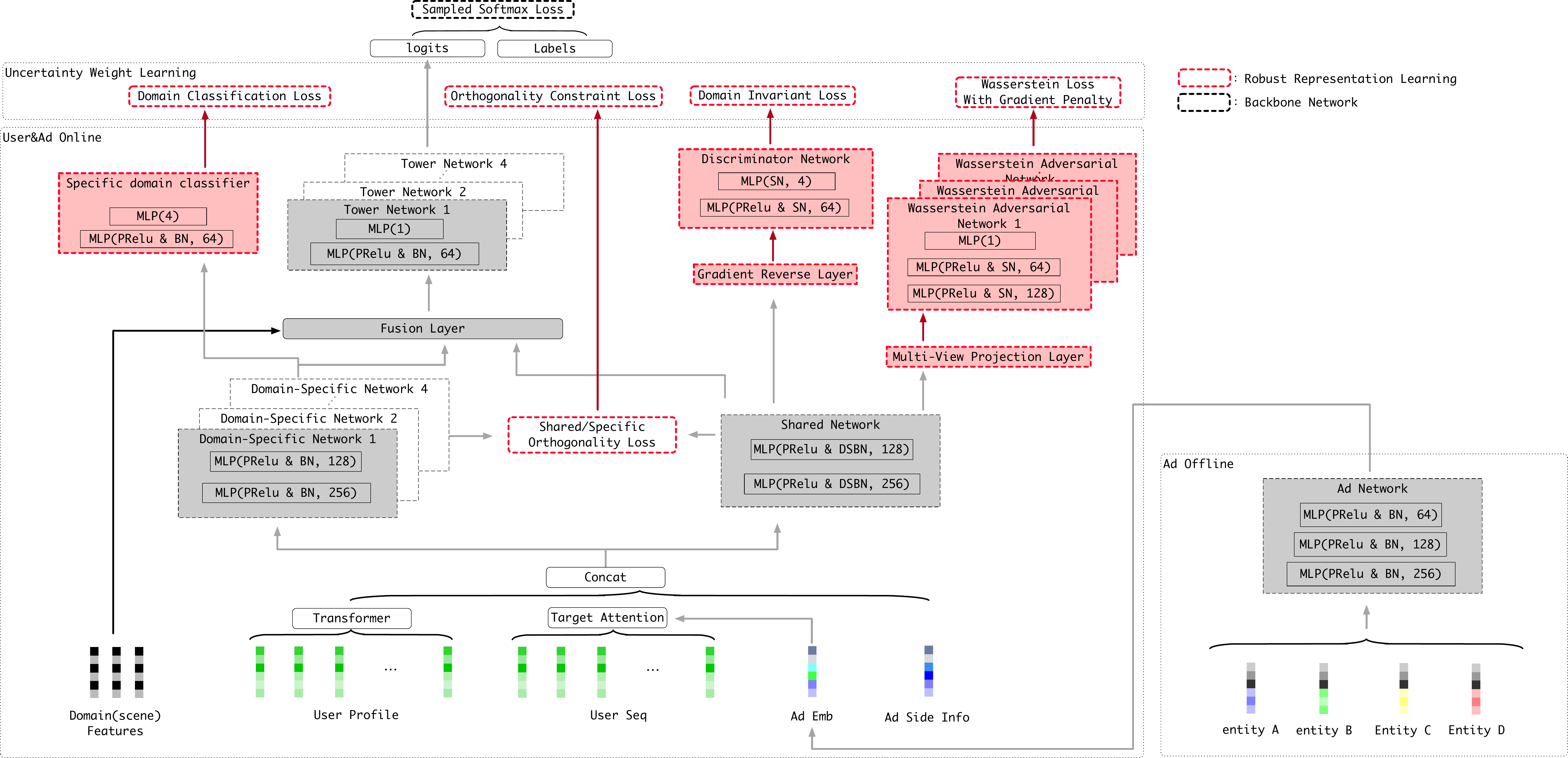}
\caption{The overall architecture of our proposed method. The multi-entity/user embedding is through the ad-network/user-transformer to get the entity/user representation. The user sequence and entity representation are through the target attention layer to obtain the attention representation. Then, all the representations are concat and fed into the backbone of the hybrid expert framework. The shared/specific-domain representations strengthen the learning capability through robust representation learning (red-boxed section). Finally, the model outputs the logits and calculates the sampled softmax loss.}
\label{fig_framework}
\end{figure*}

\section{Method}
The modeling of the unified online matching method proposed in this work focuses on the matching stage. Therefore, this section introduces the specific methods used to unify multi-domain matching modeling. The problem Formulation of unified multi-domain modeling is introduced in \ref{Formulation}. The framework of the whole matching model is introduced in \ref{Overview}. The specific implementation details of the backbone framework and RRL are introduced in \ref{BF} and \ref{RRL}, respectively.

\subsection{Problem Formulation}\label{Formulation}
The unified matching method aims to retrieve top-k multi-entity ads from a vast collection of ads by predicting users' preferences in each domain. 
Given dataset $\{\mathcal{A}, \mathcal{U}\} \in {\mathbb{R}^{P}}$, $M$ different entities set $\mathcal{A}=\{A^1,A^2,...,A^M\}$, $\mathcal{A}^m=\{a^1_m,a^2_m,...,a^T_m\}$, user set $\mathcal{U}=\{u^1,u^2,...,u^N\}$, $K$ different domains $\mathcal{D}=\{D^1,D^2,...,D^K\}$. Each domain is defined as $D^k=\{a,u\}$, including user-entity historical interaction within the domain.
A user could have different interactions with the same entities across different domains.
The RS usually learns the embeddings $\mathcal{E}_u$ of users and $\mathcal{E}_a$ of entities through neural networks. 

The matching model in RS is usually defined as a function $\mathcal{F_{\theta}}$: $\mathcal{E}_u \times  \mathcal{E}_a \to \hat{y}({a \mid u,k})$.
The goal is to get the matching scores vector $\hat{y}({a \mid u,k})$ for each user by learning user-entity information in the domain $D^k$. The formal definition is as follows
\begin{align}
\hat{y}(a^t \mid u, k)=\frac{\mathcal{F_{\theta}}(\mathcal{E}_a^t  \mid \mathcal{E}_u, \mathcal{D}^k)}{\sum_{i=0}^{|a|}(\mathcal{F_{\theta}}(\mathcal{E}_a^i  \mid \mathcal{E}_u, \mathcal{D}^k)) },
\end{align}
Here, vector $\hat{y}({a \mid u,k})$ indicates the likelihood that user $u$ is interested in the all interacted (positive) sample $a$ in the multi-entity dataset. The $a^t$ represents any kind of entity advertisement.

The vector of real values $\hat{y}({a \mid u,k})$  usually turns to a vector of probability values that sum to 1 by softmax function. However, due to the large number of interacted entities for one user, calculating the probability values for all becomes impractical. The sampled softmax function\cite{GoogleDNN} is a better way, which calculates the probability value on the sampled subset $a^\prime$ containing the target entity $a^t$. The formula is as follows:
\begin{align}
\label{sampled_softmax}
\mathcal{F_{\theta}}(\mathcal{E}_{a}  \mid \mathcal{E}_u, \mathcal{D}^k,\mathcal{Q})=
\mathcal{F_{\theta}}(\mathcal{E}_{a}  \mid \mathcal{E}_u, \mathcal{D}^k)-\log(\mathcal{Q}(a \mid u))\\
\hat{y}(a^t \mid u, k)=\frac{\mathcal{F_{\theta}}(\mathcal{E}_a^t  \mid \mathcal{E}_u, \mathcal{D}^k,\mathcal{Q})}
{\sum_{a^\prime \in a}(\mathcal{F_{\theta}}(\mathcal{E}_{a^\prime}  \mid \mathcal{E}_u, \mathcal{D}^k,\mathcal{Q})) }
\end{align}
where $\mathcal{Q}(a \mid u)$ is the sampling distribution, representing the probability of sampling class $a$ given $u$. The $a^\prime$ ($\left | a^\prime \right |  \ll \left | a \right |$) is the subset sampled by $\mathcal{Q}(a \mid u)$. 

Specifically, the objective of the matching model is to minimize the cross-entropy loss between the true label $y$ and the predicted value $\hat{y}$ computed through the sampled softmax function. The realistic significance of $y$ in the matching model is the 0/1 value of impression or not. 
The objective function can be simplified to minimize the likelihood function, as the formula:
\begin{align}
\label{loss_ce}
\min\mathcal{L}_{\theta}&=\min\sum_{k \in K}\sum_{y \in (0,1)} -{y_{k}log(\hat{y}_{k})} \nonumber  \\
&=\min{\sum_{k \in K}\sum_{(a,u) \in \mathcal{D}^k }-log(\hat{y}(a \mid u, k)
)}
\end{align}
By minimizing the $\theta$ with Eq. \ref{sampled_softmax} and Eq. \ref{loss_ce}, the matching model can make unified personalized recommendations that are relevant to the user's interests for various entities within each domain.


\begin{table}[!ht]
    \centering
\caption{FEATURE DEFINITIONS FOR THE PROPOSED METHOD.}
\begin{tabularx}{\columnwidth}{ll}
   \toprule
   Feature & Description  \\
   \midrule
   $E_{u}$ & user profiles, user-entity history/real-time feedback, etc. \\
   $E_{a}$ & entity content, entity history feedback, cross attribute, etc  \\
   $E_{seq}$ & user-entity exposure sequence  \\
   $E_{aid}$ & entity identifier attribute\\
   $E_{did}$ & multi-domain class indicator attribute\\
   $E_{ac}$ & multi-entity class indicator attribute\\
   \bottomrule
\end{tabularx}
\label{tab:feature}
\end{table}

\subsection{Architecture Overview}\label{Overview}
In this section, we introduce the overall architecture of the matching modeling process of our proposed method as shown in Figure \ref{fig_framework}. The proposed framework involves the ad network module, user transformer module, target attention module, and RRL module. The input features into the entire framework are listed in Table \ref{tab:feature}. Multi-entity features are reconstructed through the unified data link to realize unified input on the model side without additional processing.

\textit{How can the dataset comprising these features get the corresponding probability values for each user through the framework?}
In the proposed framework, instance 
is firstly fed into the shared embedding layer for dimensionality reduction, resulting in the low-dimensional embedding $\mathcal{E}=\{E_u, E_a, E_{seq}, E_{aid}, E_{did}, E_{ac}\}$. The ${E_a}$ is input into the ad network module designed by multi-layer perception (MLP) to capture entity information, extracting unified-entity embedding ${E_{uee}}$.
Simultaneously, $E_u$ is input into the user encoder module, which is a standard transformer\cite{transformer} module, to extract user representation $E_{tran}$. The $E_{tran}$ captures global information and internal dependencies within the user's behavior sequence.
Furthermore, the $E_{seq}$ and $E_{aid}$ obtain attention representation $E_{att}$ concerning the user's behavior sequence embedding using the target attention module. 
Finally, The improved representation $\mathcal{R}=\{E_{uee}, E_{tran}, E_{att}, E_{aid}\}$ is input into the backbone framework with RRL module to generate the probability value $\hat{y}$, representing the user's interest in each interacting entity.

\begin{algorithm}
\caption{Backbone Framework with RRL}
\label{algo_RRL}
\KwIn{number of domains is $K$, domain indicator embedding is $E_{did}$, extracted representation from dataset is $\mathcal{R}$, domain label is $y_{id}$, the ground-truth user-entity label is $y$.}
\KwOut{$\mathcal{L}_{final}$, $\hat{y}$}
\textbf{Init}: {Initialize $\theta^\ast$ for one shared-domain expert, 
$\{\theta_d^k\}_{k=0}^K$ for $K$ specific-domain experts, 
$\{\theta_t^k\}_{k=0}^K$ for $K$ specific-tower experts, 
$\{\theta_g^k\}_{k=0}^K$ for $K$ gate units, respectively.
}\\
\For{ $epoch=1$ {\bfseries to} $MaxEpoch$}{
    $\{\mathcal{R}^k\}_{k=0}^K = \{\mathcal{R} \cdot \mathds{1}_{\{y_{id} = k\}}\}_{k=0}^K$;\\
    \For{ $k=0$ {\bfseries to} $K$}{
        $E_{did}^k= E_{did}[k]$, $y_{id}^k= y_{id} \cdot \mathds{1}_{\{y_{id} = k\}}$; \\
        $E_{d}^k=\text{MLP}_d^k(\mathcal{R}^k|\theta^k_d)$; \\
        $E_{wd}^k=E_{s}^k=\text{MLP}_s(\mathcal{R}^k|\theta^\ast)$; \\ 
        \tcp*[h]{$\theta^\ast$ updated by all domains.}\\
        \If{is training = True}{
            $\mathcal{L}_{OCL}^k$ = OCL($E_{d}^k$, $E_{s}^k$); \\
            $\mathcal{L}_{ADL}^k$ = DAL($E_{d}^k$, $E_{s}^k$, $y_{id}^k$);
        }
        $w_g^k=\text{Sigmoid}(E_{did}^k \cdot \theta_g^k)$;\\
        $E_{d}^k$=$w_g^k[0] \cdot E_{d}^k$, $E_{s}^k$=$w_g^k[1] \cdot E_{s}^k$;\\
        $E_{g}^k=\begin{bmatrix}
                        E_{d}^k & E_{d}^k \odot E_{s}^k & E_{s}^k
                        \end{bmatrix}$;\\  
        $\hat{y}^k=\text{sampled\_softmax}(\text{MLP}_t^k(E_{g}^k|\theta^k_t))$,
        }
    \If{is training = True}{
        $E_{wd} = \begin{bmatrix}\{E_{wd}^k \cdot \mathds{1}_{y_{id} = k}\}_{k=0}^K\end{bmatrix}$;\\
        $\mathcal{L}_{wd}=\text{WDL}(E_{wd})$;
    }
    $\mathcal{L}_{ce}=\sum_{k}\sum_{i} -{y^k_ilog(\hat{y}^k_i)}$;\\
    $\mathcal{L}_{final}=\mathcal{L}_{ce}+\text{UWL}(\mathcal{L}_{wd},\mathcal{L}_{OCL},\mathcal{L}_{ADL})$;
}
\end{algorithm}

\subsection{Backbone Framework}\label{BF}
Through the modules introduced above, we obtain 
the input $\mathcal{R}=\{E_{uee}, E_{tran}, E_{att}, E_{aid}\}$ for the entire backbone framework. The detailed backbone framework combined with RRL is shown in Algorithm \ref{algo_RRL}. We introduce the implementation of the entire backbone framework in this section. The backbone framework consists of one shared-domain expert and $K$ specific-domain experts, $K$ gate units and $K$ tower experts. The process of generating matching scores from $\mathcal{R}$ can be expressed as the following formula:
\begin{align}
\hat{y}(\mathcal{R} \mid k) &= \mathcal{F}_{\theta}(\mathcal{R} \mid \theta^\ast, \theta_d^k, \theta_g^k, \theta_t^k ) \\
&=\text{MLP}_t^k(\text{Gate}^k(E_{s},E_{d}^k,E_{did}^k \mid \theta_g^k);\theta_t^k)\\
E_{s}&=\text{MLP}_s(\mathcal{R} \mid  \theta^\ast), 
E_{d}^k=\text{MLP}_d^k(\mathcal{R}^k \mid \theta^k_d)
\end{align}
where the $\mathcal{R}$ is input into the function $\mathcal{F}_\theta$, which is trained within the backbone framework.
The network structure of the $k$-th domain comprises the set $\{\text{MLP}_s, \text{MLP}_d^k, \text{Gate}^k, \text{MLP}_t^k\}$.

\subsubsection{Shared \& Specific Experts}
The $\mathcal{R}$ from all domains collaborate to train the shared-domain expert parameterized by $\theta^\ast$, which extracts the domain-invariant representation $E_{s}$ that reflects the model's generalization ability. The broader the generality of $E_{s}$, the more comprehensive the model's understanding.
Simultaneously, the $\{\mathcal{R}^k\}_{k=0}^K$ from their respective domains train the specific-domain expert parameterized by $\{\theta_d^k\}_{k=0}^K$, which capture domain-specific knowledge $\{E_{d}^k\}_{k=0}^K$. 
The greater the differences between $\{E_{d}^k\}_{k=0}^K$, the higher the recall accuracy of the model within specific domains.
While $E_{s}$ and $E_{d}^k$ capture diverse domain knowledge, the $k$-th domain may have specific requirements for them. Therefore, the backbone framework adaptively controls the degree of information interaction and fusion through the gate unit to achieve domain personalized adjustment.

\subsubsection{Gate Units}
The $k$-th gate unit learns the $k$-th domain's weight vector $w_g^k$ from the multi-domain class indicator $E_{did}^k$. The $w_g^k$ is a two-dimensional vector, indicating the importance of $E_{s}$ and $E_{d}^k$ in the $k$-th domain. The $E_{s}$ and $E_{d}^k$ are concatenated by $w_g^k$ to form a fusion representation $E_{g}^k$, the formula is as follows:
\begin{align}
w_g^k  &= \text{Sigmoid}(E_d^k \cdot \theta_g^k);\\
E_{d}^k  &= w_g^k[0] \cdot E_{d}^k, E_{s} = w_g^k[1] \cdot E_{s}; \\
E_{g}^k  &= \begin{bmatrix}
            E_{d}^k & E_{d}^k \odot E_{s} & E_{s}
            \end{bmatrix};
\end{align}
where the $\odot$ denotes hadamard-product. 
The combined representation $E_{g}^k$ is obtained by combining weighted $E_{s}$, $E_{d}^k$, and their hadamard-product fusion $E_{had}^k$. 
The direct concatenation of $E_{s}$ and $E_{d}^k$ provides a coarse-grained raw relationship between the general and specific-domain knowledge, offering a high-level understanding of the interaction.
Fine-grained relationships and co-occurrence information are captured by performing an element-wise calculation between $E_{s}$ and $E_{d}^k$, offering an underlying-level understanding of the interaction. 
The $E_{g}^k$ benefits from the different levels of granularity, providing a more comprehensive and expressive interaction representation, capturing both coarse-grained and fine-grained relationships and leveraging the complementary strengths of different representations. 

\subsubsection{specific-domain Tower Experts}
The fusion representation $E_{g}^k$ is input into $k$-th tower expert to generate the corresponding matching scores. The formula is as follows:
\begin{align}
E_t^k=\text{MLP}_t^k(E_{g}^k \mid \theta^k_t).
\end{align}
The actual value vector, denoted as $E_t^k$, is the output from the $k$-th tower expert, converted to probability $\hat{y}^k$ according to Eq. \ref{sampled_softmax}. The $\hat{y}^k$ represents the matching scores between user $u$ and the interacted entities $a$ in the $k$ domain. 
By extending Eq. \ref{loss_ce}, optimization objectives based on the backbone framework can be obtained:
\begin{align}
\label{loss_ce2}
\min\mathcal{L}_{ce}&=\min_{\theta^\ast, \theta_d^k, \theta_g^k, \theta_t^k}{\sum_k\sum_i-log(\hat{y}^{i,k})
)}
\end{align}

\subsection{Robust Representation Learning}\label{RRL}
Based on the backbone framework, RRL includes domain adversarial learning, multi-view wasserstein distribution learning, orthogonal constraint, and uncertainty weight learning. In this section, we detail the implementation of RRL.

\begin{algorithm}
\caption{Domain Adversarial Learning}
\label{algo_DAL}
\KwIn{shared/specific domain representations are $E_s$/$E_d$, shared/specific domain classifiers are $D_s/D_d$, domain label is $y_{id}$.}
\KwOut{$\mathcal{L}_s$, $\mathcal{L}_d$}
\textbf{Init}: {Initialize $\varpi_s$ for $D_s$, 
$\varpi_d$ for $D_d$.
}\\
$y_{id}$ = OneHotEncoder($y_{id}$);\\

$\hat{y}_s  = \text{softmax}(D_s(E_s|\varpi_s))$;\\
$\hat{y}_d  = \text{softmax}(D_d(E_d|\varpi_d))$;\\
${\mathcal{L}_{s}}=-\sum_i y_{id}^i \log(\hat{y_s^i})$;\\
${\mathcal{L}_{d}}=-\sum_i y_{id}^i \log(\hat{y_d^i}$;\\

\BlankLine
\tcp*[h]{Update of domain experts.}\\
$\theta^{\ast} \leftarrow {\theta^{\ast}} - \mu \bigtriangledown_{\theta^{\ast}},
\theta_{d} \leftarrow {\theta_{d}} - \mu \bigtriangledown_{\theta_{d}}$;\\
\BlankLine
\tcp*[h]{Update of domain classifiers.}\\
${\varpi_s} \leftarrow {\varpi_s}+\mu \bigtriangledown_{\varpi_s}, 
{\varpi_d} \leftarrow {\varpi_d}-\mu \bigtriangledown_{\varpi_d}$;\\

\end{algorithm}

\subsubsection{Domain Adversarial Learning}
Through the backbone framework of hybrid experts, we get the ideal matching score. However, in the above process, we assume that the shared-domain expert and specific-domain experts can provide good learning representations.
In fact, the entire framework simply distinguishes the capabilities of different components by assigning different initialization parameters to the various experts. Much research has proven it is a problem. 
Therefore, we use domain adversarial learning (DAL) constraints $E_{s}$ to learn generalized knowledge and strengthen the recognition of $E_{d}$ for unique domain information. The specific structure of DAL is shown in algorithm \ref{algo_DAL}.

Generative adversarial learning\cite{GAN} is a method to promote the generator to generate more realistic samples through the minimax game between the generator and discriminator. We learn from adversarial domain adaptive learning and adopt the gradient reversal layer\cite{DANN} to realize the adversarial process of shared-domain expert $E_s$(generator) and shared-domain classifier $V_s$(discriminator). 
By introducing adversarial loss, the $V_s$ no longer accurately reflects the domain the sample belongs to, forcing the $E_s$ to learn more domain-general knowledge. The formula for domain invariant loss is as follows:
\begin{align}
\text{min} {\mathcal{L}_{s}}=
\text{min}_{\theta^\ast}\text{max}_{\varpi_s}-\sum_i^{\left | y_{id} \right |} \sum_k^K y_{id}^{i,k} \log(\hat{y_s}^{i,k}),\\
\hat{y}_s^{i,k} = \text{softmax}(V_s^{i,k})= \frac{e^{V_s^{i,k}}}{\sum_{j}^K e^{V_s^{i,j}}}, 
\end{align}
where the ${\varpi_s}$ is the parameter of $V_s$, the $\hat{y}_s$ is the probability value reflecting the domain that $E_s$ belongs to, and the $y_{id}$ is the ground-truth label composed of the domain indicator $\{k=0, 1, ..., K\}$ encoded as the one-hot vector. 
The optimization of the adversarial loss involves a minimax optimization problem, where ${\varpi_s}$ updates along the gradient minimization direction while $\theta^\ast$ updates along the gradient maximization direction. The more general knowledge the $E_s$ learns, the more difficult the $V_s$ gets the correct label.
During the training process, ${\varpi_s}$ and $\theta^\ast$ find the Nash equilibrium state in the competition.

Furthermore, the addition of the spectral normalization \cite{SN} (SN) layer to the shared-domain classifier network addresses issues such as mode collapse and training instability that can occur during the adversarial training process. SN is an effective weight regularization technique that stabilizes the training process by constraining the spectral norm of the weight matrix. The formula for SN is as follows:
\begin{align}
\tilde{v} = \frac{{\varpi_s}^\top u}{|{\varpi_s}^\top u|_2},
\tilde{u} = \frac{{\varpi_s}v}{|{\varpi_s}v|_2}, \hat{\varpi_s} = \frac{{\varpi_s}}{\tilde{u}^\top {\varpi_s}\tilde{v}}
\end{align}
where the $v,u$ is randomly initialized and updated by the power iteration method to obtain the approximate singular vector $\tilde{v},\tilde{u}$.

For specific-domain experts, the $E_d$ enhances unique domain information learning through specific-domain classifier $V_d$. The loss function is shown as follows:
\begin{align}
\text{min} {\mathcal{L}_{d}}=
\text{min}_{\theta_d,\varpi_d}-\sum_i^{\left | y_{id} \right |} \sum_k^K y_{id}^{i,k} \log(\hat{y_d}^{i,k}),\\
\hat{y}_d^{i,k} = \text{softmax}(V_d^{i,k})= \frac{e^{V_d^{i,k}}}{\sum_{j}^K e^{V_d^{i,j}}}, 
\end{align}
where the ${\varpi_d}$ is the parameter of $V_d$. With the progress of training, $\theta_d$ and ${\varpi_d}$ update along the gradient minimization direction, and $V_d$ predicts the domain that $E_d$ belongs to more accurately, which means that the unique domain information in $E_d$ is constantly enhanced.

\begin{algorithm}
\caption{Multi-View Wasserstein Distribution Learning}
\label{algo_WDL}
\KwIn{shared-domain representation is $E_s$, multi-view layers are $\{P^t\}_0^T$, wasserstein critics are $\{C^t\}_0^{T-1}$.}
\KwOut{$\mathcal{L}_{wd}$}
\textbf{Init}: Initialize $\{\varsigma_p^{t}\}_{0}^{T}$ for $\{P^t\}_0^{T}$, $\{\varsigma_c^{t}\}_{0}^{T}$ for $\{C^t\}_0^{T-1}$.\\
\For{ $t=0$ {\bfseries to} $T-1$}{
    $E_p^t=P^t(E_s|\varsigma_p^t)$;
    $E_p^{t+1}=P^{t+1}(E_s|\varsigma_p^{t+1})$;\\
    $\widehat{E_p}^t=\xi E_p^t+(1-\xi)E_p^{t+1}, \text{vector }\xi\sim U(0,1)$;\\
    $\mathcal{L}_{gp}^t = \sum_i( \left\| \nabla_\mathrm{\hat{E_p^t}} C^{t}(\hat{E_p^t}) \right\|_2 - 1)^2$;\\
    $E_c^t=C^t(E_p^t|\varsigma_c^t)$;
    $E_c^{t+1}=C^{t}(E_p^{t+1}|\varsigma_c^{t})$;\\
    $s^t = \text{weight}(E_c^t)$;
    $s^{t+1} = \text{weight}(E_c^{t+1})$; \\
    $\mathcal{L}_{wd}^t = \frac{1}{\left | \mathcal{D}   \right | }({\textstyle \sum_i{s^{t+1,i}}}-{\textstyle \sum_i{s^{t,i}}}+\mathcal{L}_{gp}^t )$;\\
}
$\mathcal{L}_{wd}=\sum_t^{T-1}\mathcal{L}_{wd}^t$;\\
\BlankLine
\tcp*[h]{Update of multi-view layers.}\\
${\varsigma_p^{t}} \leftarrow {\varsigma_p^{t}}-\mu \bigtriangledown_{\varsigma_p^{t}}$; 
${\varsigma_p^{t+1}} \leftarrow {\varsigma_p^{t+1}}-\mu \bigtriangledown_{\varsigma_p^{t+1}}$;\\
\tcp*[h]{Update of wasserstein critics.}\\
 ${\varsigma_c^{t}} \leftarrow {\varsigma_c^{t}}+\mu \bigtriangledown_{\varsigma_c^{t}}$; 
\end{algorithm}

\subsubsection{Multi-View Wasserstein Distribution Learning}
The shared-domain classifier aims to learn $E_s$ to classify into the incorrect domain. However, this goal allows $E_s$ to obtain domain-invariant and wrong confounding knowledge. To further constrain the learning of $E_s$ and reduce the wrong direction guidance, aligning the representation distribution of different domains directly in the representation space is reasonable. 
Indeed, the wasserstein distance is a metric that can measure the distance between two probability distributions smoothly and continuously, even when they have no overlap. This property provides meaningful gradient information during optimization stably. 
Therefore, we also design Multi-View Wasserstein Distribution Learning(MVWDL) for the shared-domain expert in addition to DAL. MVWDL limits misleading guidance for $E_s$ and aids in learning more generalized knowledge by aligning distributions across domains in the multi-view representation space. The specific structure of MVWDL is shown in algorithm \ref{algo_WDL}.

When the model serves $K$ domains, MVWDL consists of projection layers $\{P^t\}_0^T$ ($T=K$) and wasserstein critics $\{C^t\}_0^{T-1}$.
By learning distinct feature mappings, each projection layer $P^t$ captures a domain-specific perspective $E_p^t=\{E_p^{t,k}\}_{k=0}^{K}$ from the domain-invariant representation $E_s$, which is derived from all domain data. 
$\{E_p^t\}_0^T$ is a collection of domain-specific projections, where $E_p^{t,k}$ indicates that it belongs to the $k$-th domain and is in the $t$-th domain perspective projection space. The $t$-th multi-view projection layer is defined as follows:
\begin{align}
E_p^t=P^t(E_s|\varsigma_p^t);
\label{weighted}
\end{align}

All projection representations are pairwise grouped into $\{E_p^t,E_p^{t+1}\}_{t=0}^{T-1}$. MVWDL does $T-1$ distribution alignment by wasserstein critics $\{C^t\}_0^{T-1}$ for them.
The $C^t$ takes $E_p^t$ and $E_p^{t+1}$ as input respectively and outputs the score vector $E_c^t$ and $E_c^{t+1}$.
$E_c^t$ is a collection of score values, $E_c^t=\{E_c^{t,k}\}_{k=0}^{K}$, representing the actual values and not normalized probability values. The $t$-th wasserstein critic is defined as follows:
\begin{align}
E_c^t&=C^t(E_p^t|\varsigma_c^t);\\
s^t &= \text{weight}(E_c^t)=\begin{cases}
\frac{1}{\left | \mathcal{D}^k \right | }E_c^{t,k}   & \text{ if } t=k \\
 \frac{1}{\left | \mathcal{D} \right | -\left | \mathcal{D}^k \right | }E_c^{t,k}  & \text{ if } t\ne k
\end{cases}
;
\label{weighted}
\end{align}
In the MVWDL, the shared representation input into each projection layer comes from multiple domains. So, the $E_p^{t,k}$ belonging to the $k$-th domain is projected into the domain perspective projection space of both its own domain ($t=k$) and other domains ($t\ne k$).
To handle this scenario, we design a weighted method to adjust the scores appropriately. The weight for each score value $E_c^{t,k}$ is determined based on the amount of data in the respective domains, as shown in Eq.\ref{weighted}. 
If $E_c^{t,k}$ belongs to $k$-th domain and also belongs to $t$ domain perspective projection space where $t=k$, 
it is multiplied by $\frac{1}{\left | \mathcal{D}^k \right | }$, which represents the amount of data in the t-th domain. 
On the other hand, if $E_c^{t,k}$ belongs to $k$-th domain but is projected into a $t$-th domain perspective projection space where
$t\ne k$, it is multiplied by $\frac{1}{\left | \mathcal{D} \right | -\left | \mathcal{D}^k \right | }$, which represents the total amount of data in other domains. 

The optimization goal of MVWDL is to minimize the wasserstein distance between distributions represented by different projections.
The scores obtained by the critics are weighted according to the prior domain knowledge, highlighting the importance of the projection representations belonging to different domains from the same domain perspective when calculating the wasserstein distance.
It helps to determine the training priorities of different regions of data distribution in the projection Spaces. The loss function of MVWDL is shown as follows:
\begin{align}
\label{Lwd}
\text{min }\mathcal{L}_{wd}& = \sum_t^{T-1}\text{min }\mathcal{L}_{wd}^t;\\
\text{min }\mathcal{L}_{wd}^t & = \text{min}_{\theta^\ast,\varsigma_p^{t}, \varsigma_p^{t+1}}
\text{max}_{\varsigma_c^{t}}
{\sum_i^{\left | \mathcal{D} \right | }}s^{t+1,i}-{\sum_i^{\left | \mathcal{D} \right | }}s^{t,i} \nonumber 
\\&+\sum_i^{\left | \mathcal{D} \right | }( \left\| \nabla_\mathrm{\hat{E_p^t}} C^t(\hat{E_c^t}) \right\|_2 - 1)^2
;\\
\widehat{E_p}^t=\xi E_p^t&+(1-\xi)E_p^{t+1}, \text{vector }\xi\sim U(0,1);
\end{align}
The optimization of $\mathcal{L}_{wd}^t$ is the minimax problem. The $\{\theta^\ast, \varsigma_p^{t}, \varsigma_p^{t+1}\}$ are updated along the gradient minimization direction, while the $\varsigma_c^{t}$ is updated along the gradient maximization direction. The better the shared representation, the smaller the wasserstein distance between pairwise projection representations and the more difficult the wasserstein critic distinguishes.
Minimizing $\mathcal{L}_{wd}^t$ means minimizing the score difference between the distribution of the $t$ domain perspective projection and the $t+1$ domain perspective projection. 
$\widehat{E_p}^t$ is the linear interpolated set of $E_p^t$ and $E_p^{t+1}$, controlled by a random scalar vector $\xi$ sampled from a uniform distribution.
Inspired by WGAN-GP\cite{WGAN-GP}, the gradient penalty ensures $C^t$ satisfies the 1-Lipschitz condition by ensuring the gradient of the $C^t$ concerning the interpolated samples has a norm close to 1.

\subsubsection{Orthogonality Constraints Learning}
An orthogonality constraint loss is introduced to address the model's limitation in separating the domain-invariant representation space from the specific-domain representation space. This loss function aims to penalize redundant information in these spaces and encourage shared-domain and specific-domain experts to learn distinct knowledge. The orthogonality constraint loss is defined as follows:
\begin{align}
\mathcal{L}_{\text{ortho}} =  {\textstyle \sum_{k=0}^{K}} \left\| {E_s^k}^\top E_d^k  - \mathbf{I} \right\|_\text{F}^2;
\end{align}
Here, ${E_s^k}^\top$ represents the transpose of the domain-invariant representation ${E_s^k}$, $\mathbf{I}$ is the identity matrix, and $|\cdot|_{\text{F}}$ denotes the Frobenius norm.
The objective of the orthogonality loss is to encourage the domain-invariant representation ${E_s^k}$ and the specific-domain representation ${E_d^k}$ to be orthogonal to each other in the embedding space. This is achieved by minimizing the difference between the product of ${E_s^k}^\top$ and ${E_d^k}$ and the identity matrix $\mathbf{I}$.

By minimizing the orthogonality loss, the model promotes the learned matrices to possess better orthogonality properties, ensuring that the shared-domain and specific-domain representations capture distinct and non-redundant knowledge. This can ultimately improve the performance and generalization ability of the model.
\subsubsection{Uncertainty Weight Learning for Losses}
The optimization objective of the proposed method incorporates multiple loss functions, including ADL, MVWDL, and OCL. Each of these loss functions contributes to improving different aspects of the overall method. However, determining the weights of these auxiliary loss functions manually or based on empirical parameters can be inefficient and challenging. Additionally, as the optimization progresses, the importance of each auxiliary loss may change.
To address this issue, we draw inspiration from Uncertainty Weighted Loss (UWL)\cite{UWL} and propose setting the weights of different loss functions by considering the uncertain homoscedasticity of each auxiliary loss. The final loss function is represented as follows:
\begin{align}
\mathcal{L}_{\text{final}}(\sigma) = \mathcal{L}_{ce} + \sum_{i=0}^{3} (\frac{1}{2\sigma_i^2}\mathcal{L}_{i} + \log{\sigma_i});
\end{align}

The $\mathcal{L}_{\text{final}}$ denotes the final loss function, $\{\mathcal{L}_i\}_{i=0}^{3}$ represents the auxiliary loss functions $\{\mathcal{L}_{s}, \mathcal{L}_{d}, \mathcal{L}_{wd}, \mathcal{L}_{ortho} \}$. 
The parameter $\sigma$ is the training parameter during the model training process, representing the relative difficulty between different auxiliary losses. As the value of $\sigma$ increases, it indicates higher uncertainty and noise in the output of the corresponding auxiliary task. In order to account for this difficulty, the weight assigned to the loss decreases. This adjustment allows for more effective model training by appropriately balancing the contributions of each auxiliary task.

By considering the homoscedasticity uncertainty and adjusting the weights accordingly, the UWL method ensures that each task contributes meaningfully to the training process. This approach leads to improved performance and robustness of the method across different domains, as it considers the auxiliary tasks' varying difficulty levels.

\begin{figure}
\centering
\includegraphics[scale=0.5]{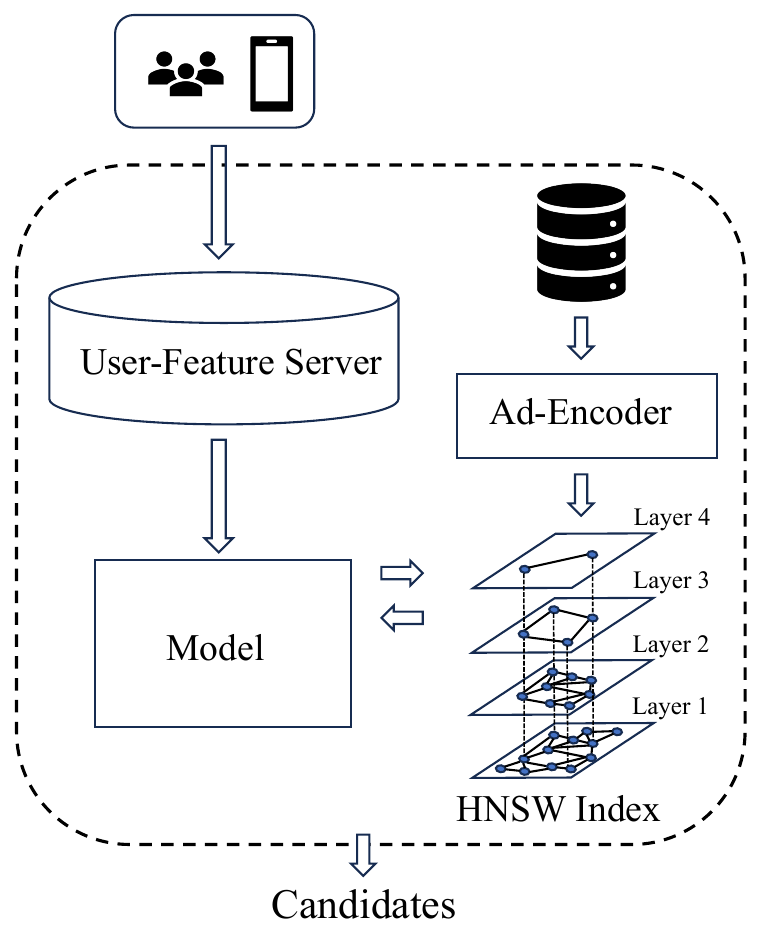}
\caption{The overall architecture of the unified matching service in online recommendation system.}
\label{fig_RS}
\end{figure}

\section{Online Matching System}
In this section, we mainly introduce the overall data link and some implementation details of the unified online matching system.
\subsection{Unified Data Link}
Ads from different entities often have significant differences in features, magnitudes, and presentation forms, and are often independent and isolated in data links. Therefore, the first step is to strive for consistency on the input side of the underlying features.
To ensure the best possible alignment of different entity features, we adopted a unified feature system when constructing multi-entity features. This means that we strive to align and integrate the characteristics of various entities as much as possible. For instance, when dealing with live-streaming sale ads, we obtain the product that corresponds to the live-streaming sale and utilize the features of the product to complete the live-broadcast advertisement. However, it is crucial to recognize that different advertising types may possess their own distinct features. In such cases, we fill in the default values for these unique features that may be missing from other entities. It allows us to achieve a unified modeling approach in our system, known as Unified Modeling in Entity Space (UMES).
By processing in this way, the data, features, and modeling methods of multiple entities can be unified, reducing operation and maintenance costs.
\subsection{Top-K Retrieval Based On HNSW}
The currently commonly used structure in the industry is the twin-tower structure: during training, user representations and item representations are independently learned, and then the inner product of the two representations is used; during the online inferring, the top-k items that are nearest neighbors to the user representation are retrieved through ANN\cite{ADI}. The reason for this is that the scoring set faced by retrieval is tens of millions of levels, while the sorting stage usually has hundreds to thousands of candidate sets. However, the method of compressing the user/item representation and calculating the inner product severely limits the user-item interaction ability and the model expression ability (for example, the attention of the historical interaction sequence of the user and target item cannot be calculated); therefore, we continue to use HNSW\cite{hnsw} in NANN\cite{nann} Graph retrieval technology: During training, we learn user and item representations, calculate attention, and use multi-layer DNN to obtain the final scoring result. Then, we use the L2 distance of the items to construct a 4-layer HNSW graph structure. In the last online service, we use model scoring and HNSW's greedy search algorithm to select the top-n results with the highest scores in each layer, and finally merge these top candidate sets and select the top-k ones with the highest scores as the output of the retrieval module. By constructing the HNSW graph index structure and implementing the greedy retrieval algorithm, we can introduce complex model structures while reducing the number of scoring candidate sets to upgrade the model's expressive capabilities.

\subsection{Matching As A Service}
Eventually, our matching model will be encapsulated into service in the advertising engine and provide services for online traffic in the form of \textsc{\textit{user in, topk-ad out}}. After the top-k candidates are retrieved, they will be further sent to the ranking, re-ranking, and other recommender system modules, as shown in Figure \ref{fig_RS}.

\begin{table}[h]
    \centering
    \caption{THE STATISTIC INFORMATION OF THE ALIBABA DISPLAY ADVERTISING DATASET.}
    \label{tab:dataset}
    \begin{tabular}{lllll}
        \toprule
        Statistics & Dataset     & Domain     & Record        &  \\
        \midrule
        Users      & 124,253,906 & Domain \#1 & 499,036,447   &  \\
        Entity \#A & 5,562,826   & Domain \#2 & 466,070,990   &  \\
        Entity \#B & 85,825      & Domain \#3 & 606,781,383   &  \\
        Entity \#C & 8,130       & Domain \#4 & 1,093,375,741 &  \\
        Entity \#D & 74,447      & Domain \#all & 2,665,264,561 & \\
\bottomrule
\end{tabular}
\end{table}



\section{Experiments}
\subsection{Experimental Setup}
\subsubsection{Datasets}
Since there is no public dataset of multi-entity data, we use real industrial data, namely Alibaba Display Advertising Data. The Table \ref{tab:dataset} shows the statistical information of this dataset.
For model training and testing, a total of actual log data of users in four domains is collected with 124,253,906 users. For each user, each impressed item is marked as a positive example, and unimpressed items are recorded as negative examples, resulting in a total of 2,665,264,561 records.

\subsubsection{Compared Models}
\begin{compactitem}
\item DNN\cite{yi2019sampling}: This method learns mixed data from different domains based on a shared model.
\item Shared-Bottom\cite{shared_bottom}: A classic multi-domain model that shares a shared layer at the bottom and builds different towers based on different domain data.
\item Cross-Stitch\cite{cross}: This method introduces cross-stitch units to combine different domains' hidden layers with learned linear weights.
\item MMoE\cite{mtlMMoE}: A multi-gated hybrid-expert method that assigns input features to multiple expert networks. It weighs and fuses the outputs of different experts through the gated network. 
\item Dselectk-MoE\cite{dselectk}: This variant of MoE introduces the novel DSelect-k, a continuously differentiable sparse gate for MoE. It provides explicit control over the number of expert selections.
\item PLE\cite{ple}: It refines the hybrid-expert into two types: shared-domain and specific-domain expert. PlE strengthens the learning of these two expert types through the selective combination of the input of each domain.
\item Backbone-RRL: This is the best version of our proposed method. It combines the backbone framework with the ADL, MVWDL, OCL, and UWL modules.
\end{compactitem}

\begin{table*}
    \centering
\caption{OVERALL PERFORMANCE COMPARISONS ON ALIBABA DISPLAY DATASET IN RECALL-ALL@2000 METRICS FOR 2000 USERS.}
\resizebox{\linewidth}{!}{
\begin{tabular}{lllllllllll}
   \toprule
\multicolumn{1}{l}{\multirow{2}{*}{Method}} & \multicolumn{2}{c}{Domain \#1}     & \multicolumn{2}{c}{Domain \#2}    & \multicolumn{2}{c}{Domain \#3}     & \multicolumn{2}{c}{Domain \#4}     & \multicolumn{2}{c}{Average}   \\
   \cline{2-11}
\multicolumn{1}{c}{} & Recall-all & Relative Change  & Recall-all & Relative Change & Recall-all & Relative Change  & Recall-all & Relative Change  & Recall-all & Relative Change  \\
   \midrule
DNN & 0.3420 & -  & 0.3840 & - & 0.5599 & -  & 0.2671 & -  & 0.3882 & -  \\
Shared-Bottom    & 0.3549 & 3.77\%  & 0.4038 & 5.16\% & 0.5753 & 2.75\%  & 0.3034 & 13.59\% & 0.4093 & 5.44\%  \\
Cross-Stitch     & 0.3504 & 2.46\%  & 0.3963    & 3.20\% & 0.5715 & 2.07\%  & 0.2906 & 8.80\%  & 0.4022 & 3.61\%  \\
MMoE    & 0.3546 & 3.68\%  & 0.3926 & 2.24\% & 0.5789 & 3.39\%  & 0.3047 & 14.08\% & 0.4077 & 5.02\%  \\
DselectK-MoE     & 0.3648 & 6.67\%  & 0.4009 & 4.40\% & 0.5730 & 2.34\%  & 0.2926 & 9.55\%  & 0.4078 & 5.05\%  \\
PLE & 0.3596 & 5.15\%  & 0.4084 & 6.35\% & 0.5911 & 5.57\%  & 0.3183 & 19.17\% & 0.4193 & 8.01\%  \\
\textbf{Backbone\&RRL}    & \textbf{0.3927} & \textbf{14.82\%} & \textbf{0.4200} & \textbf{9.37\%} & \textbf{0.6180} & \textbf{10.38\%} & \textbf{0.3187} & \textbf{19.32\%} & \textbf{0.4373} & \textbf{12.65\%} \\
   \bottomrule
\end{tabular}
}
\label{tab:Recall-all}
\end{table*}

\begin{figure}
\centering
\includegraphics[scale=0.4]{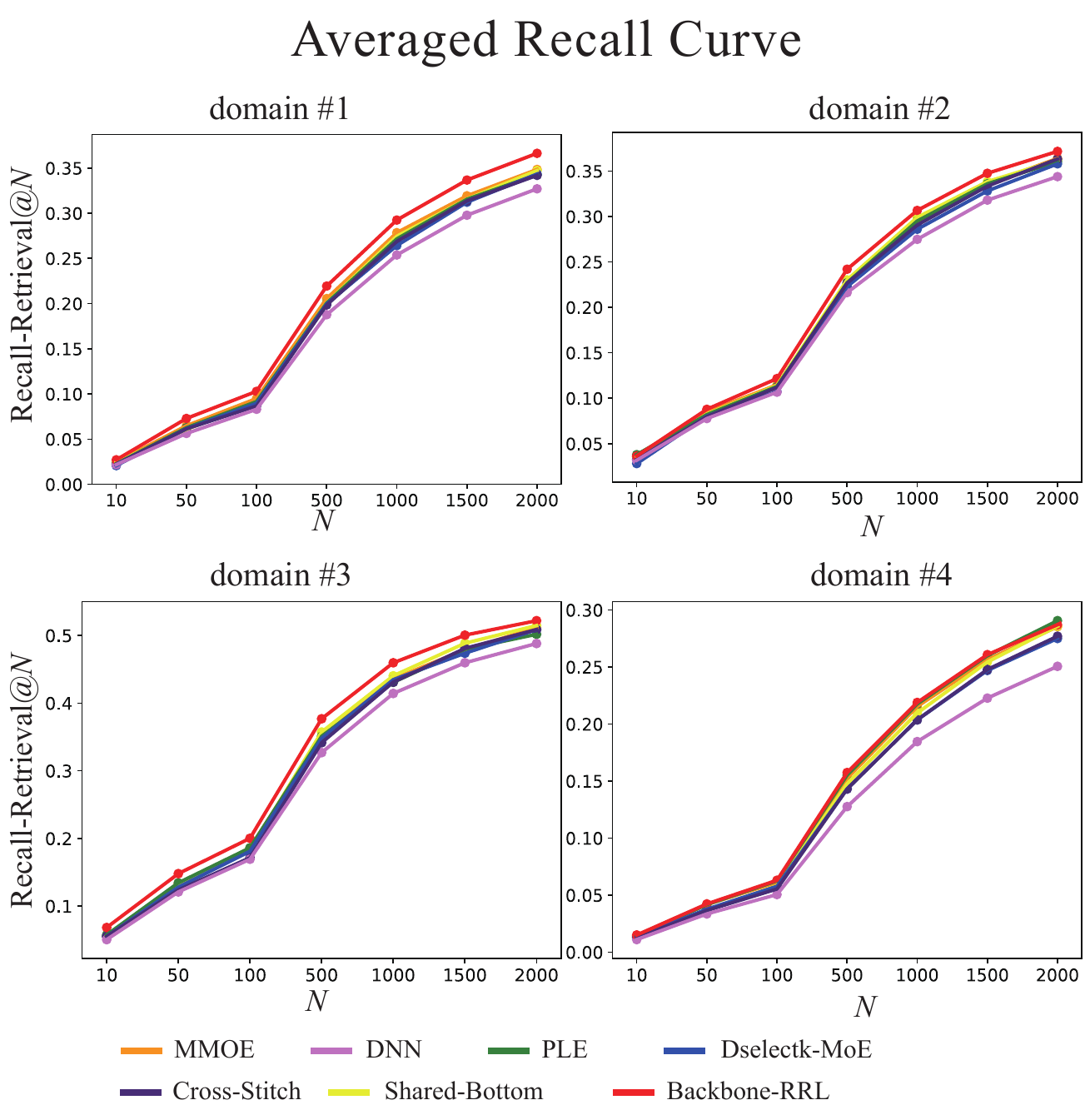}
\caption{Average recall-retrieval curves of Backbone-RRL and other methods on Alibaba display dataset in different numbers of retrieved entities for 2000 users.}
\label{fig_recall}
\end{figure}
\begin{table*}[htbp]
    \centering
\caption{THE PERFORMANCE OF THE PROPOSED METHOD IN THE ABLATION EXPERIMENTS. THE `RELATIVE CHANGE' REPRESENTS THE RELATIVE IMPROVEMENT IN RECALL-RETRIEVAL@2000 METRICS FOR 2000 USERS.}
\begin{tabular}{llllll}
   \toprule
    & Domain \#1  & Domain \#2 &  Domain \#3  & Domain \#4 & Average \\
  Method&  Relative Change &  Relative Change & Relative Change & Relative Change  & Relative Change\\
   \midrule
   Backbone  & -  & - & -  & - &-\\
   Backbone \& DAL   & 5.45\%  &1.96\%  &-0.29\%  &3.37\% &2.62\% \\
   Backbone \& DAL \& UWL   &	5.11\%  &3.20\% 	 &1.13\%  &3.78\% &3.30\% \\
   Backbone \& MVWDL  &6.50\%  &1.31\%   & 0.19\% &2.33\% &2.58\%\\
   \textbf{Backbone \& RRL}    &\textbf{7.00\%}   & \textbf{4.83\%}  &\textbf{1.25\%}	 &\textbf{7.30\%} &\textbf{5.10\%}\\
   \bottomrule
\end{tabular}
\label{tab:ablation}
\end{table*}

\subsubsection{Implementation Details}
The six comparison methods and the five versions of our proposed method differ only in the design of the backbone and RRL module. However, the other modules, including the unified data link, ad network module, and user encoder module, are all configured in the same manner.
During feature processing, we convert all sparse features into dense embeddings and concatenate them with dense features. All methods are trained and tested using the same framework, with consistent optimization settings such as the optimizer, learning rate, and batch size.
The backbone frameworks for all methods are 4-layer MLPs, with each hidden layer having a dimension of $[256, 128, 64, 1]$. The DAL module is a single-layer MLP with a dimension of 64, while MVWDL is a 3-layer MLP with dimensions of $[128, 64, 1]$. In the version without UWL configuration, the weight coefficients of all auxiliary losses are fixed at $0.01$.

It is important to note that the RRL modules are only used during the training phase. These modules assist the backbone framework in generating more robust representations, thereby improving overall performance. However, once training is complete, these modules have already influenced the parameters for generating matching scores, and no additional calculations are required.
Considering factors such as computational complexity and inference time during online usage, it becomes more practical to deliver only the parameters of the backbone framework without the additional modules. This approach reduces the computational burden during online inference, ensuring faster response times and improved efficiency.
\subsubsection{Evaluation and Metrics}
To evaluate the effectiveness of different methods, we use the Recall-all@$N$ and Recall-retrieval@$N$ metrics\cite{nann}. 
The Recall-all@$N$ metric is used to assess methods by exhaustively evaluating the entities for all users. The Recall-retrieval@$N$ metric is used to assess methods by traversing the graph-based index to retrieve relevant entities for all users. They are defined as the formula:
\begin{align}
\text{Recall-all}(\mathcal{A}_u,\mathcal{G}_u)@N=\frac{ \left |\mathcal{G}_u  \cap  \mathcal{A}_u \right | }{ \left |\mathcal{G}_u   \right | }, \\
\text{Recall-retrieval}(\mathcal{R}_u,\mathcal{G}_u)@N=\frac{ \left |\mathcal{G}_u  \cap  \mathcal{R}_u \right | }{ \left |\mathcal{G}_u   \right | },
\end{align}
The set of ground truth entities for user $u$ is defined as $\mathcal{G}_u$. The $\mathcal{A}_u$($\left |\mathcal{A}_u   \right |=N$) is the set of top-$N$ scored entities for a user $u$ that can be produced by brute-force scanning. The $\mathcal{R}_u$($\left |\mathcal{R}_u   \right |=N$) is the set of relevant entities retrieved by the graph-based index for a user $u$. 

\subsection{Offline Evaluation and Results}
The evaluation of the overall performance of the seven methods on the test set of the Alibaba display advertising dataset is presented in Table \ref{tab:Recall-all} and Figure \ref{fig_recall}. From table \ref{tab:Recall-all}, we can see that our proposed Backbone\&RRL method achieves a higher average recall-all metric across all domains compared to the other methods. When considering the recall-all metric for each individual domain, Backbone\&RRL also demonstrates significant advantages. Although PLE performs similarly to Backbone\&RRL in Domain \#4, it performs much worse than Backbone\&RRL in other domains. Figure \ref{fig_recall} illustrates the average recall-retrieval curves of all methods on the Alibaba display advertising dataset for different numbers of retrieval entities. When the number of retrievals is small ($N \in [0,100]$), Backbone\&RRL maintains its leading advantage in all domains, indicating that the method can recommend entity ads that users prefer at the forefront of the recall queue. As the number of retrievals increases, Backbone\&RRL continues to exhibit significant advantages in domains \#1, \#2, and \#3. However, in domains \#4, PLE gradually demonstrates similar performance to Backbone\&RRL. Although It can gradually catch up with the performance of Backbone\&RRL in domains \#4, it also implies that it requires more retrieval entities, resulting in less efficiency.

\subsection{Ablation Study}
In this section, we conducted experiments to evaluate the impact of each module in the RRL on the multi-domain matching model. 
\begin{compactitem}
\item Backbone: The overall design of the Backbone structure of our method is the same as that of PLE. The advancement of the Backbone structure is to use the \textbf{CONCAT}-version gated unit\cite{ADI}.
\item Backbone-DAL: This is one version of our method. The backbone framework combines the ADL module and the OCL module. The weight coefficients of all auxiliary losses are set by empirical parameters.
\item Backbone-MVWDL: This is one version of our method. The backbone framework combines the MVWDL module and the OCL module. The weight coefficients of all auxiliary losses are set by empirical parameters.
\item Backbone-DAL\&UWL: This is one version of our method. The backbone framework combines the ADL module and the OCL module. The weight coefficients of all auxiliary losses are automatically optimized by the UWL module during training.
\end{compactitem}

We kept the backbone framework and OCL module fixed while varying the other modules in five different combinations for training and testing. The results are presented in Table \ref{tab:ablation}.
From the results, we can observe that compared to the Backbone model with only the backbone framework, the models Bachbone\&DAL and Bachbone\&MVWDL show positive improvements in the overall recall performance. However, Bachbone\&MVWDL performs better in domain \#1 and domain \#3, while Bachbone\&DAL performs better in domain \#2 and domain \#4. This indicates that the seesaw effect is still apparent, and different modules also have varying emphases on different domains. Furthermore, the introduction of the UWL module in the Bachbone\&DAL\&UWL model significantly enhances the recall performance in domain \#2 and domain \#4 compared to the Bachbone\&DAL module. This demonstrates that the dynamic update weight mechanism of the UWL model can effectively alleviate the seesaw effect, as opposed to a fixed weight coefficient. Finally, the Bachbone\&RRL model, which integrates all modules, exhibits the best recall performance across all domains.

\subsection{Visualization}

\begin{figure}
\centering
\includegraphics[scale=0.28]{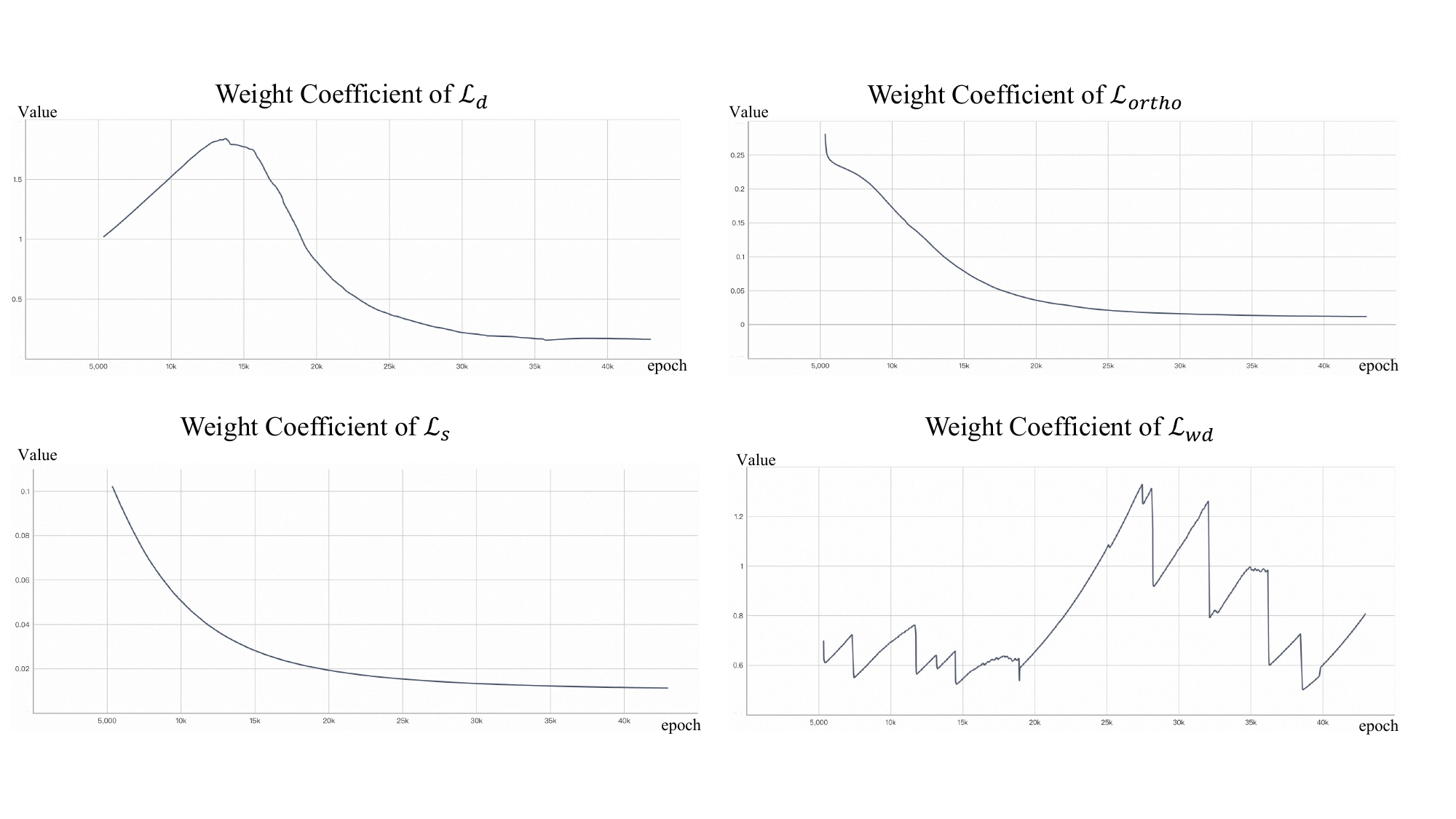}
\caption{Visualize the change of weight coefficients of four auxiliary losses during the training process.}
\label{fig_factor}
\end{figure}

\subsubsection{Uncertainty Weight Learning for Losses}
In Figure \ref{fig_factor}, we visualize the optimization curves of the weight coefficients of four auxiliary losses during the training processes.
The changing trends of the four weight coefficients are quite different. The weight coefficients of $\mathcal{L}_{s}$ and $\mathcal{L}_{ortho}$ have been decreasing until they level off. It indicates that their uncertainty and noise are higher at the beginning. Thus, UWL reduces the weight coefficients to decrease their effect on the overall training. The weight coefficient of $\mathcal{L}_{d}$ first increases and then decreases and finally levels off, indicating that the early specific-domain experts have higher constraint certainty and have a more apparent positive effect on the overall training. The weight coefficient of $\mathcal{L}_{wd}$ tends to oscillate continuously, indicating that the impact of aligning the domain-invariant representation distribution through MVWDL display varies significantly during the training process, with higher uncertainty and more training difficulty, so that UWL adjusts its weight coefficient oscillatingly. The result shows that it is not advisable to pre-set fixed empirical parameters because the impact of auxiliary loss and training difficulty will continue to change with the training process. UWL can effectively improve the impact of uncertainty in the training process on model performance.

\begin{figure}[htbp]
\centering
\includegraphics[scale=0.28]{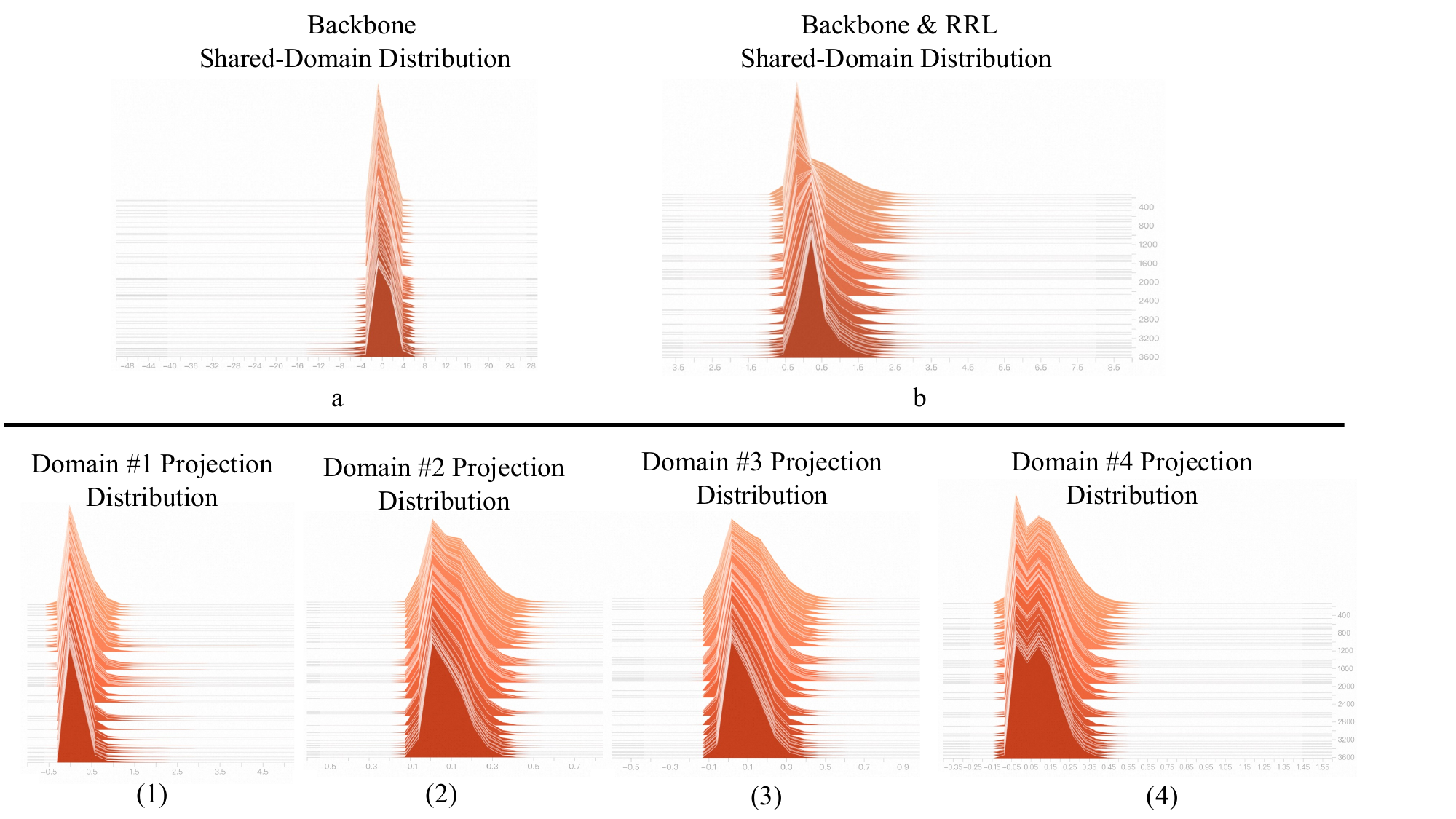}
\caption{The first line is the distribution of shared-domain representation from the Backbone model and Backbone\&RRL model. The second line is the distribution of shared-domain projection representation under four domains by the Backbone\&RRL model. }
\label{fig_distribution}
\end{figure}
\subsubsection{Distribution of domain-invariant representation}
We verify the effectiveness of RRL by visualizing the distribution of domain-invariant representations of the Bachbone model and Bachbone\&RRL model, as shown in Figure \ref{fig_distribution}.
In the Backbone model, since there are no additional constraints and guided design, the distribution of domain-invariant representations does not change significantly during the iteration process. It means the model cannot effectively adjust and optimize the distribution of shared representations. In contrast, in the Bachbone\&RRL model, the distribution of shared representations changes evidently as iterations proceed. Guided by WDL and RRL, the model can better capture and learn the commonalities of the data to better adapt to the characteristics of the data.
Further, we can observe the projected representation distribution of domain-invariant representations guided by MVWDL in four domains. From Figure \ref{fig_distribution}.(1)-(4), It can be seen that the distribution states of these four projection representations are different, which is related to the data distribution in the four domains. The representation distribution obtained through shared representation projection can be guided to obtain a representation distribution with simple domain characteristics by Eq.\ref{weighted}. The effect of this correct projection stems from the domain-invariant knowledge of the shared representation distribution in the Bachbone\&RRL model, which can exclude confusing and misleading information.

\subsection{Online A/B Test}
We have deployed our model on the display advertising system of Alibaba. To get a stable conclusion, we observed the online experiment for about a week. Two common metrics in the advertising system are used to measure online performance: RPM~(Revenue Per Mille), and Cost. As shown in Table \ref{tab:online_domains}, the present method gets overall improvements on four domains in our online A/B test experiment.
\begin{table}[htbp]
\caption{ONLINE A/B TEST IN DIFFERENT DOMAINS}
\label{tab:online_domains}
\scalebox{0.9}{
    \begin{tabular}{c c c c c}
    \toprule
    Online Metric &  Domain \#1 & Domain \#2 & Domain \#3 & Domain \#4  \\
    \midrule
    COST & +0.3\% & +0.0\% & +0.7\% & +1.9\%  \\
    RPM & +0.5\% & +0.1\% & +1.0\% & +2.0\%\\ 
    \bottomrule
    \end{tabular}
}
\end{table}

Meanwhile, since we have designed data links and data solutions for the unified modeling of multiple entities, the problem of unfair scoring of different entities by the model should be solved. As is shown in Table \ref{tab:online_entities}, we observed the online effects of the online experiment on different entities for a week. We can see that the PV~(Page View) and COST of the entities \#B, \#C, and \#D have all been greatly improved. This experiment proves that under the unified modeling of multiple entities in the entire advertising system, traffic is transferred from the dominant entity \#A to other entities. This effectively alleviates the problem of unfair scoring of different types of ads by the RS.

\begin{table}[htbp]
\caption{ONLINE A/B TEST ON DIFFERENT ENTITIES}
\label{tab:online_entities}
\scalebox{1.0}{
    \begin{tabular}{c c c c c}
    \toprule
    Online Metric &  Entity \#A & Entity \#B & Entity \#C & Entity \#D  \\
    \midrule
    PV & -0.8\% & +29.8\% & +81.3\% & +64.3\%\\ 
    COST & -0.3\% & +19.2\% & +30.0\% & +63.5\%  \\
    \bottomrule
    \end{tabular}
}
\end{table}

\section{Conclusion}
In this paper, we propose a new unified online matching method with modeling of multi-entity and multi-domain. By constructing unified modeling in entity space, this system ensures the consistency of online/offline training and the fairness of online multi-entity retrieval. Furthermore, The RRL, as a plug-and-play configuration for any backbone framework module, strengthens the learning capabilities of shared/specific representations from multiple perspectives. Our experiments on the industrial dataset verify the superiority of the proposed method. The function of each module has been verified through ablation experiments. The unified matching method we proposed has been deployed on the Alibaba display advertising system and has achieved considerable profits, verifying the commercial value of the proposed method.

\bibliographystyle{IEEEtran}
\bibliography{ref.bib}
\end{document}